\documentclass[english,aps, pra, amsmath,amssymb, reprint]{revtex4-1}
\usepackage[T1]{fontenc}
\usepackage[utf8]{inputenc}
\usepackage{mathtools}
\usepackage{amsmath}
\usepackage{amssymb}
\usepackage{esint}

\makeatletter
\usepackage{slashed}

\makeatother

\usepackage{babel}
\begin{document}

\title{Grassmann phase-space methods for fermions: uncovering classical
probability structure}

\author{Evgeny A. Polyakov}
\email{e.a.polyakov@gmail.com}

\affiliation{Russian Quantum Center, Novaya 100, 143025 Skolkovo, Moskow Region,
Russia }
\begin{abstract}
The phase-space description of bosonic quantum systems has numerous
applications in such fields as quantum optics, trapped ultracold atoms,
and transport phenomena. Extension of this description to the case
of fermionic systems leads to formal Grassmann phase-space quasiprobability
distributions and master equations. The latter are usually considered
as not possessing probabillistic interpretation and as not directly
computationally accessible. Here, we describe how to construct $c$-number
interpretations of Grassmann phase-space representations and their
master equations. As a specific example, the Grassmann $B$ representation
is considered. We disscuss how to introduce $c$-number probability
distributions on Grassmann algebra and how to integrate them. A measure
of size and proximity is defined for Grassmann numbers, and the Grassmann
derivatives are introduced which are based on infinitesimal variations
of function arguments. An example of $c$-number interpretation of
formal Grassmann equations is presented.
\end{abstract}


\maketitle

\section{INTRODUCTION}

Phase-space approach to quantum mechanics has proved to be invaluable
tool in such fields as quantum optics and trapped ultracold atoms
\citep{Ng2011,Deuar2007,Deuar2011,He2012}. This approach allows one
to calculate quantum observable properties as averages of classical
quantities over certain quasiprobability distributions. At the same
time, the full quantum evolution often takes the form of a simple
Fokker-Planck equation for these quasiprobability distributions. The
latter property turns phase-space techniques into a stochastic simulation
tool which was used to conduct Monte Carlo calculations of a number
of full many-body problems \citep{Ng2011,Deuar2007,Deuar2011,He2012}.

When extending phase-space techniques to the case of fermions, a fundamental
limitation is faced due to anticommutation of fermionic canonical
variables. Because of this, the corresponding canonical operators
cannot have $c$-number eigenvalues except zero. As a consequence
it is impossible to construct $c$ number quasiprobability distributions
for them. 

There are several workarounds to this problem \citep{Corney2004,Corney2006,Corney2006b}.
The most formal and non-classical one is to change the notion of number
\citep{Cahill1999}. Fermionic canonical operators can have non-zero
eigenvalues if we consider these eigenvalues as anticommuting numbers
which are conventionally called Grassmann numbers (hereinafter, the
term ``Grassmann number'' will be abbreviated as ``$g$ number'').
This way it is possible to develop phase-space representations for
fermions which bear remarkable analogy to bosonic ones. In particular,
there are Grassmann quasiprobability distributions of the same types:
$P$, $B$, $Q$ functions, $s$-ordered representations \citep{Cahill1999,Plimak2001,Dalton2016},
and also Wigner functions \citep{Mrowczynski2013}. More over, their
master equations also look quite similar to the bosonic case. For
example, in the case of real-time quantum dynamics with pairwise interactions,
it is possible to derive master equation which looks similar to the
Fokker-Planck equation for positive-$P$ distribution \citep{Plimak2001,Dalton2016}. 

Nevertheless, there is important difference: all fermionic quasiprobability
distributions are $g$ numbers. Grassmann numbers are dramatically
different from $c$ numbers: the latter are simple, the most basic
things. However $g$ number is not simple: it has the structure of
many-body correlated state. Every $g$ number defines a hierarchy
of $n$-point functions, just as the physical state defines a hierarchy
of correlations.

Because of this complexity, Grassmann phase space methods are usually
considered as not possessing probabillistic interpretation and as
not directly computationally accessible \citep{Corney2006b,Corney2006,Davidson2016}.
At the same time, there are published works in which $c$ number stochastic
unravelings are constructed for Grassmann master equations \citep{Plimak2001,Dalton2016}.
These findings rise a number of questions. Firstly, the possibility
of stochastic unraveling means that the Grassmann representations
are in fact equivalent to certain $c$ number quasiprobability distributions,
with their own correspondence rules for observables, for quantum states,
and for evolution equations. These equivalent $c$ number quasiprobabilities
were not considered in the literature. Secondly, in the work \citep{Dalton2016}
there exists controversy with the earlier paper \citep{Plimak2001}.
This means that the nature of this stochastic unraveling is not completely
understood.

The goal of this work is to clarify these questions. In fact, these
questions have fundamental dimension: they imply that whatever formal
\textit{one-time} representation of quantum mechanics we invent, with
respect to classical computability there are only $c$-number distributions,
and nothing more. Such a unified view opens the way for general theorems
to be formulated and proven. For example, generalized phase-space
methods solve the sign problem \citep{Deuar2002}: they allow us to
represent quantum evolution as a stochastic process which can be simulated
by Monte Carlo methods. Despite this success, in all practical realizations
of simulation protocols the problem of quantum complexity reappears
in one or another form e.g. numerical instability or exponential spread
of Monte Carlo trajectories. Is that a fundamental limitation or just
imperfection of our knowledge?

Another field for which these questions are relevant, is the derivation
and $c$-number stochastic unraveling of master equations for open
quantum systems in fermionic environments \citep{Shi2013,Chen2013}. 

For the purpose of this work, we choose a particular $g$-number phase-space
method, the Grassmann $B$ representation \citep{Dalton2016}. This
representation is analog of Drummond-Gardiner positive-$P$ representation
\citep{Drummond1980}. This specific choice does not reduce the generality
of results: had we chosen another $g$-number representation, we would
apply to it the same techniques as described in this work, and would
come to analogous conclusions.

In section \ref{sec:GRASSMANN-B-REPRESENTATION} we begin with a brief
exposition of Grassmann $B$ representation as it is known in literature
and simultaneously recalling basic facts about Grassmann calculus.
Next, we are going to construct probability and stochastic calculus
on Grassmann algebra. In order to accomplish this, in section \ref{sec:GRASSMANN-CALCULUS-REVISITED}
we discuss such notions as: function of arbitrary $g$ number; proximity
and size of $g$ number; Grassmann derivatives based on infinitesimal
variation of function argument; probability distribution on Grassmann
algebra and its integral. In section \ref{sec:STOCHASTIC-GRASSMANN-B-REPRESENTATION}
we introduce $c$-number probability distribution into the Grassmann
$B$ representation. Actually, this way we obtain a novel $c$ number
representation which we call ``stochastic Grassmann $B$ representation''.
It is shown that evolution of the emerging quasiprobability distributions
is governed by Fokker-Planck equation (for systems with pairwise interactions).
The corresponding stochastic equations are found to coincide with
that derived in \citep{Dalton2016}. We conclude in section \ref{sec:CONCLUSION}.

\section{GRASSMANN $B$ REPRESENTATION\label{sec:GRASSMANN-B-REPRESENTATION}}

The $g$-number analog of Drummond-Gardiner positive-$P$ representation
\citep{Drummond1980} is Grassmann $B$ representation. In this section
we briefly review main results about $B$ representation, simultaneously
recalling basic notions of Grassmann calculus.

\subsection{Definition of representation}

Suppose we have a fermionic system with $M$ modes (single-particle
states). For each mode $j$, there are associated creation $\widehat{a}_{j}^{\dagger}$
and annihilation $\widehat{a}_{j}$ operators. A Bargmann coherent
state is defined as 
\begin{equation}
\left|\boldsymbol{e}\right\rangle =\exp\left(-\sum_{j}e_{j}\widehat{a}_{j}^{\dagger}\right)\left|0\right\rangle ,
\end{equation}
where $e_{j}$ is a Grassmann number. We consider $g$ numbers $e_{j}$
as linearly independent basis elements, with anticommuting multiplication
law 
\begin{equation}
e_{i}e_{j}=-e_{j}e_{i},
\end{equation}
and which generate the algebra of arbitrary $g$ numbers 
\begin{multline}
g=G\left(0\right)+\sum_{i}G\left(i\right)e_{i}+\sum_{i_{1}<i_{2}}G\left(i_{1}i_{2}\right)e_{i_{1}}e_{i_{2}}\\
\ldots+\sum_{i_{1}<\ldots<i_{M}}G\left(i_{1}\ldots i_{M}\right)e_{i_{1}}\ldots e_{i_{M}}.\label{eq:grassmann_as_n_point_hierarchy-1}
\end{multline}
Every Grassmann number $g$ can be unambiguously decomposed into even
and odd parts,
\begin{equation}
g=g^{+}+g^{-},
\end{equation}
where the even part $g^{+}$ consists of even powers of $e_{i}$ in
representation Eq. (\ref{eq:grassmann_as_n_point_hierarchy-1}), and
the odd part $g^{-}$ consists of odd powers of $e_{i}$ respectively.
We also suppose that there is Grassmann complex conjugation operation
\begin{equation}
\left(\right)^{*}:\,e_{j}\to e_{j}^{*},
\end{equation}
which is analog of complex conjugation for $c$ numbers. The elements
$e_{j}^{*}$ generate arbitrary conjugated $g$ numbers $g^{*}$,
a conjugated variant of (\ref{eq:grassmann_as_n_point_hierarchy-1}).
 Although the most general $g$ number contains both elements $e_{i}$
and their conjugates $e_{j}^{*}$, we do not encounter such $g$ numbers
in our problem, thus we assume that all $g$ numbers contain either
$e_{j}$ or $e_{j}^{*}$. To put it in other way, we are dealing only
with ``analytic'' $g$ numbers. We assume that complex conjugation
respects anticommutativity: for any $g$ numbers $\alpha$, $\beta$,
$\gamma$ we have: 
\begin{equation}
\left(\alpha\beta\gamma\right)^{*}=\gamma^{*}\beta^{*}\alpha^{*}.\label{eq:product_complex_conjugation}
\end{equation}
Due to this rule, Grassmann complex conjugation can be interpreted
in a way which is consistent with Hermitian conjugation. For example:
\begin{equation}
\left(\alpha\widehat{a}_{j}\gamma\right)^{\dagger}=\gamma^{*}\widehat{a}_{j}^{\dagger}\alpha^{*}.\label{eq:product_Hermitian_conjugation}
\end{equation}
 Annihilation and creation operators act upon coherent state as:
\begin{equation}
\widehat{a}_{j}\left|\boldsymbol{e}\right\rangle =e_{j}\left|\boldsymbol{e}\right\rangle ,\,\,\,\,\widehat{a}_{j}^{\dagger}\left|\boldsymbol{e}\right\rangle =-\overrightarrow{\partial}_{j}\left|\boldsymbol{e}\right\rangle =\left|\boldsymbol{e}\right\rangle \overleftarrow{\partial}_{j},\label{eq:coherent_state_properties}
\end{equation}
where $\overrightarrow{\partial}_{j}$ and $\overleftarrow{\partial}_{j}$
are the usual left and right Grassmann derivative operators with respect
to element $e_{j}$. Left and right derivatives with respect to complex
conjugate elements $e_{j}^{*}$ are denoted as $\overrightarrow{\partial}_{j^{*}}$
and $\overleftarrow{\partial}_{j^{*}}$. In order to maintain consistency
with the properties (\ref{eq:product_complex_conjugation}) and (\ref{eq:product_Hermitian_conjugation}),
the complex conjugation of derivatives is defined as 
\begin{equation}
\left[\overrightarrow{\partial}_{j}\right]^{*}=\overleftarrow{\partial}_{j^{*}},\,\,\,\,\left[\overleftarrow{\partial}_{j}\right]^{*}=\overrightarrow{\partial}_{j^{*}}.
\end{equation}
For example:
\begin{equation}
\left(\alpha\overrightarrow{\partial}_{j}\beta\gamma\right)^{*}=\gamma^{*}\beta^{*}\overleftarrow{\partial}_{j^{*}}\alpha^{*}
\end{equation}
and
\begin{equation}
\left(\alpha\widehat{a}_{j}\gamma\overrightarrow{\partial}_{j}\right)^{\dagger}=\overleftarrow{\partial}_{j^{*}}\gamma^{*}\widehat{a}_{j}^{\dagger}\alpha^{*}.
\end{equation}
 Analogously to bosonic Drummond-Gardiner positive-$P$ representation
\citep{Drummond1980}, we can double the dimension of our grassmann
algebra by introducing additional basis elements $e_{j}^{\prime}$,
$j=1\ldots M$. Then, one introduces the non-diagonal coherent state
projections $\left|\boldsymbol{e}\right\rangle \left\langle \boldsymbol{e}^{\prime*}\right|$
, where $\left\langle \boldsymbol{e}^{\prime*}\right|=\left(\left|\boldsymbol{e}^{\prime}\right\rangle \right)^{\dagger}=\left\langle 0\right|\exp\left(-\sum_{j}\widehat{a}_{j}e_{j}^{\prime*}\right)$.
Every number-conserving density operator can be expanded over these
projections as
\begin{equation}
\widehat{\rho}=\int de_{1}^{\prime*}\ldots de_{M}^{\prime*}de_{M}\ldots de_{1}B\left(\boldsymbol{e},\boldsymbol{e}^{\prime*}\right)\left|\boldsymbol{e}\right\rangle \left\langle \boldsymbol{e}^{\prime*}\right|,\label{eq:B_representation}
\end{equation}
where $\int de_{j}$ is a standard Grassmann integration; $B\left(\boldsymbol{e},\boldsymbol{e}^{\prime*}\right)$
is even $g$ number which is called $B$ representation of density
operator. We will denote the relation (\ref{eq:B_representation})
symbolically as 
\begin{equation}
B\left(\boldsymbol{e},\boldsymbol{e}^{\prime*}\right)=\left\{ \widehat{\rho}\right\} _{B}.
\end{equation}
In work \citep{Dalton2016} it is shown that $B\left(\boldsymbol{e},\boldsymbol{e}^{\prime*}\right)$
always exists and is unique.

\subsection{Equation of motion \label{subsec:grassmann_B_equation_of_motion}}

Let us consider a quantum system with Hamiltonian
\begin{equation}
\widehat{H}=\widehat{a}_{p}^{\dagger}T_{pq}\widehat{a}_{q}-\frac{1}{4}\widehat{a}_{p}^{\dagger}\widehat{a}_{q}^{\dagger}V_{pqrs}\widehat{a}_{r}\widehat{a}_{s}.\label{eq:standard_hamiltonian-1}
\end{equation}
From now on, summation over repeated indices is implied. Real time
evolution of density operator is governed by von Neumann equation
\begin{equation}
i\partial_{t}\widehat{\rho}=\left[\widehat{H},\widehat{\rho}\right].\label{eq:von_neumann_equation-1}
\end{equation}
We can use the properties of coherent states (\ref{eq:coherent_state_properties})
in order to find master equation for the corresponding $B$ representation.
In particular, by Grassmann integration by parts it can be shown that
\citep{Dalton2016} 
\begin{equation}
\left\{ \widehat{a}_{j}\widehat{\rho}\right\} _{B}=e_{j}\left\{ \widehat{\rho}\right\} _{B},\,\,\,\,\left\{ \widehat{a}_{j}^{\dagger}\widehat{\rho}\right\} _{B}=\overrightarrow{\partial}_{j}\left\{ \widehat{\rho}\right\} _{B},
\end{equation}
\begin{equation}
\left\{ \widehat{\rho}\widehat{a}_{j}\right\} _{B}=\left\{ \widehat{\rho}\right\} _{B}\overleftarrow{\partial}_{j^{*}}^{\prime},\,\,\,\,\left\{ \widehat{\rho}\widehat{a}_{j}^{\dagger}\right\} _{B}=\left\{ \widehat{\rho}\right\} _{B}e_{j}^{\prime*}.
\end{equation}
Here, $\overleftarrow{\partial}_{j^{*}}^{\prime}$ is the right Grassmann
derivative with respect to element $e_{j}^{\prime*}$. We can apply
these rules for von Neumann equation (\ref{eq:von_neumann_equation-1}),
and find:
\begin{multline}
\partial_{t}\left\{ \widehat{\rho}\right\} _{B}=-\overrightarrow{\partial}_{p}\left(iT_{pq}e_{q}\right)B-\left\{ \widehat{\rho}\right\} _{B}\left[\overrightarrow{\partial}_{p}^{\prime}\left(iT_{pq}e_{q}^{\prime}\right)\right]^{*}\\
+\frac{1}{2}\overrightarrow{\partial}_{p}\overrightarrow{\partial}_{q}\left(\frac{i}{2}V_{pqrs}e_{r}e_{s}\right)\left\{ \widehat{\rho}\right\} _{B}\\
+\frac{1}{2}\left\{ \widehat{\rho}\right\} _{B}\left[\overrightarrow{\partial}_{p}^{\prime}\overrightarrow{\partial}_{q}^{\prime}\left(\frac{i}{2}V_{pqrs}e_{r}^{\prime}e_{s}^{\prime}\right)\right]^{*}.\label{eq:B_representation_master_equation}
\end{multline}
Now, if we compare this equation with the classical probability $c$-number
Fokker-Planck equation, expressed in terms of complex variables \citep{Polyakov2015},
\begin{multline}
\partial_{t}P=-\partial_{p}A_{p}P-\partial_{p}^{*}A_{p}^{*}P+\frac{1}{2}\partial_{p}\partial_{q}B_{pl}B_{ql}P\\
+\partial_{p}\partial_{q}^{*}B_{pl}B_{ql}^{*}P+\frac{1}{2}\partial_{p}^{*}\partial_{q}^{*}B_{pl}^{*}B_{ql}^{*}P,\label{eq:classical_fokker_planck}
\end{multline}
we observe that $B$ representation master equation (\ref{eq:B_representation_master_equation})
looks like anticommuting analog of Fokker-Planck equation (\ref{eq:classical_fokker_planck}).
This analogy encourages us to find Grassmann stochastic process which
has $\left\{ \widehat{\rho}\right\} _{B}$ as its ``probability''
density. It fact, it has been done in \citep{Dalton2016}, but without
considering the emerging probability distributions. In the following
sections we will do it by explicitly introducing $c$-number probability
distribution into the $B$ representation (\ref{eq:B_representation}). 

\section{GRASSMANN CALCULUS REVISITED\label{sec:GRASSMANN-CALCULUS-REVISITED}}

We want to construct a classical stochastic interpretation of the
$B$ master equation (\ref{eq:B_representation_master_equation}).
Before we do it, we need to carry out some preparatory work. The classical
stochastic process is defined through infinitesimal increments of
the process variables. The appearance of term ``infinitesimal''
means we need to discuss how to introduce the norm of arbitrary $g$
number $g$. Moreover, the behaviour under infinitesimal variations
is described in terms of derivatives. However, the conventional Grassmann
derivative operators $\overrightarrow{\partial}_{j}$ and $\overleftarrow{\partial}_{j}$
are defined as formal algebraic manipulations on the basis elements
$e_{j}$. Therefore, we need to find Grassmann derivatives which are
connected with infinitesimal variations. Next, in order to introduce
probability distributions on Grassmann algebra, we need to discuss
the notion of function of arbitrary $g$ number and how to integrate
it.

\subsection{Norm of Grassmann number}

In {[}\citealp{Dalton2016}, p. 49{]} it is argued that $g$ numbers
do not have notions of size and mangnitude, thus there is no notion
of proximity for them. Nevertheless, we believe that this is not correct. 

Since ``analytic'' Grassmann numbers are defined according to Eq.
(\ref{eq:grassmann_as_n_point_hierarchy-1}), we see that each $g$
number is equivalent to a hierarchy of $n$-point functions $G\left(i_{1}\ldots i_{n}\right)$.
Physically, we can interpret $g$ number as a quantum many-body state,
and the functions $G\left(i_{1}\ldots i_{n}\right)$ can be interpreted
as its $n$-particle amplitudes. Due to anticommutation between the
basis elements, $G\left(i_{1}\ldots i_{n}\right)$ are not unique:
we can represent $n$-point function as a sum
\begin{equation}
G\left(i_{1}\ldots i_{n}\right)=G_{A}\left(i_{1}\ldots i_{n}\right)+Z\left(i_{1}\ldots i_{n}\right),
\end{equation}
where $G_{A}\left(i_{1}\ldots i_{n}\right)$ is comletely antisymmetric,
and $Z\left(i_{1}\ldots i_{n}\right)$ is arbitrary but which has
the symmtery of any Young tableau except complete antisymmetry. We
can introduce the norm of $g$ number as the sum of norms of its $n$-particle
amplitudes
\begin{equation}
\left\Vert g\right\Vert ^{2}\equiv\left|G_{A}\left(0\right)\right|^{2}+\left\Vert G_{A}\left(1\right)\right\Vert ^{2}+\ldots+\left\Vert G_{A}\left(M\right)\right\Vert ^{2}.\label{eq:grassmann_norm-1}
\end{equation}
Then, the distance between two $g$-numbers $g$ and $h$ is defined
as $\left\Vert g-h\right\Vert ^{2}$. Such a definition is appealing
from physical point of view, since the two quantum states should be
regarded as similar if all their $n$-point functions (correlations)
are similar. If we choose $n$-point-function norm as the Hilbert-Schmidt
norm,
\begin{equation}
\left\Vert G_{A}\left(n\right)\right\Vert ^{2}=\sum_{i_{1}<\ldots<i_{n}}\left|G_{A}\left(i_{1}\ldots i_{n}\right)\right|^{2},\label{eq:n_point_function_norm-1}
\end{equation}
then our $g$ number norm satisfies all the expected and reasonable
inequalities,
\begin{equation}
\left\Vert g+h\right\Vert \leq\left\Vert g\right\Vert +\left\Vert h\right\Vert ,\,\,\,\,\left\Vert gh\right\Vert \leq\left\Vert g\right\Vert \left\Vert h\right\Vert ,
\end{equation}
and if 
\begin{equation}
\left\Vert g-h\right\Vert =0\,\,\,\,\textrm{then}\,\,\,\,g=h.
\end{equation}
From a physical point of view, the norm (\ref{eq:n_point_function_norm-1})
has the meaning of (unnormalized) probability of observing any $n$-point
configuration, and $\left\Vert g\right\Vert ^{2}$ is its normalization
factor.

\subsection{Functions of Grassmann numbers}

\subsubsection{Algebraic functions}

The major objects of our theory, $B$ function $B\left(\boldsymbol{e},\boldsymbol{e}^{\prime*}\right)$,
coherent state dyadic $\left|\boldsymbol{e}\right\rangle \left\langle \boldsymbol{e}^{\prime*}\right|$,
and master equation (\ref{eq:B_representation_master_equation}),
are formulated as depending on basis elements $e_{j}$ and $e_{j}^{\prime*}$.
This means that in a stochastic interpretation, $e_{j}$ and $e_{j}^{\prime*}$
should be replaced with stochastic process variables $g_{j}$ and
$g_{j}^{\prime*}$, which should be considered as arbitrary $g$ numbers.
Therefore, we need to consider functions of arbitrary grassmann numbers,
e.g. $\left|\boldsymbol{g}\right\rangle \left\langle \boldsymbol{g}^{\prime*}\right|$.
General analytic function $f$ of arbirtary $g$ numbers $g_{j}$
is a sum of monomials 
\begin{equation}
\left(g_{i_{1}}^{+}\right)^{p_{1}}\ldots\left(g_{i_{n}}^{+}\right)^{p_{n}}g_{j_{1}}^{-}\ldots g_{j_{m}}^{-},
\end{equation}
where $p_{k}$ are nonnegative integer powers since in general $\left(g_{k}^{+}\right)^{2}\neq0$;
however the indices $j_{1}\ldots j_{m}$ should all be different since
$\left(g_{k}^{-}\right)^{2}=0$. We call such functions algebraic
since they can be expressed in terms of algebraic operations: multiplication,
addition, and taking even/odd parts.

\subsubsection{Non-algebraic functions. }

In order to define $c$-number stochastic process, we also need to
introduce classical probabilities on Grassmann numbers. Apparently
they cannot be expressed in terms of algebraic operations. However,
such non-algebraic functions of grassmann numbers naturally depend
on $n$-point functions. For a given $g$ number $g$, we will denote
its $n$-point function by the corresponding capital letter, $G\left(\boldsymbol{i}_{n}\right)$,
where $\boldsymbol{i}_{n}=\left(i_{1}\ldots i_{n}\right)$; the set
of all $G\left(\boldsymbol{i}_{n}\right)$ of a given order $n$,
for all values of $\boldsymbol{i}_{n}$, will be denoted by $G\left(n\right)$;
and the full hierarchy $\left(G\left(0\right),\ldots G\left(M\right)\right)$
will be designated by $G$. Therefore, a classical probability $P$
depending on $g$ will be denoted as $P\left(G_{A}\right)$. Observe
that we take antisymmetric part of $G$. 

\subsection{Metric Grassmann derivatives}

Now we have the notion of proximity and magnitude. We can introduce
the Grassmann derivatives which are based on infinitesimal variations
of arguments. In order to distinguish them from the ordinary formal
Grassmann derivatives, we call them ``metric Grasmann dervatives''.
According to standard calculus, the derivative of function $f$ is
defined through its local behaviour 
\begin{equation}
f\left(g+\delta\right)-f\left(g\right)=\sum_{\boldsymbol{i}_{n}}\Delta_{A}\left(\boldsymbol{i}_{n}\right)\partial_{G_{A}\left(\boldsymbol{i}_{n}\right)}f\left(g\right)+O\left(\left\Vert \delta\right\Vert ^{2}\right),\label{eq:c_number_calculus_local_behaviour-1}
\end{equation}
where $\Delta_{A}\left(\boldsymbol{i}_{n}\right)$ is antisymmetric
part of $n$-point function $\Delta\left(\boldsymbol{i}_{n}\right)$
of $\delta$. This definition is precise. However, it is insuficient
since it ignores the algebraic structure and the commutation properties
of $\delta$. This is because we can write 
\begin{equation}
\delta=\delta^{+}+\delta^{-},
\end{equation}
and substitute it into $f\left(g+\delta\right)$. Since $f\left(g+\delta^{+}+\delta^{-}\right)$
is a polynomial, we expand it, and move all $\delta^{+}$ and $\delta^{-}$
to the left (or to the right) respecting their commutation properties.
Keeping only the first order terms in $\delta^{+}$ and $\delta^{-}$,
we arrive at the following representations of local behaviour:
\begin{equation}
f\left(g+\delta\right)-f\left(g\right)=\delta^{+}\overrightarrow{\partial}_{g}^{+}f\left(g\right)+\delta^{-}\overrightarrow{\partial}_{g}^{-}f\left(g\right)+O\left(\left\Vert \delta\right\Vert ^{2}\right)
\end{equation}
and
\begin{equation}
f\left(g+\delta\right)-f\left(g\right)=f\left(g\right)\overleftarrow{\partial}_{g}^{+}\delta^{+}+f\left(g\right)\overleftarrow{\partial}_{g}^{-}\delta^{-}+O\left(\left\Vert \delta\right\Vert ^{2}\right),
\end{equation}
where we introduce the odd left $\overrightarrow{\partial}_{g}^{-}$,
the odd right $\overleftarrow{\partial}_{g}^{-}$, the even left $\overrightarrow{\partial}_{g}^{+}$,
and the even right $\overleftarrow{\partial}_{g}^{+}$ metric Grassmann
derivatives. The left even Grassmann derivative $\overrightarrow{\partial}_{g_{i}}^{+}$
has the following properties:
\begin{equation}
\overrightarrow{\partial}_{g_{i}}^{+}c=0,\,\,\,\,\overrightarrow{\partial}_{g_{i}}^{+}g_{j}^{+}=\delta_{ij},\,\,\,\,\overrightarrow{\partial}_{g_{i}}^{+}g_{j}^{-}=0,
\end{equation}
where $c$ is a $g$ number constant. More complex objects are differentiated
according to linearity,
\begin{multline}
\overrightarrow{\partial}_{g}^{+}\left\{ c_{1}f_{1}\left(g\right)+c_{2}f_{2}\left(g\right)\right\} \\
=c_{1}\overrightarrow{\partial}_{g}^{+}f_{1}\left(g\right)+c_{2}\overrightarrow{\partial}_{g}^{+}f_{2}\left(g\right),
\end{multline}
and by employing the following commutation relation: 
\begin{equation}
\overrightarrow{\partial}_{g}^{+}f=\left(\overrightarrow{\partial}_{g}^{+}f\right)+f\overrightarrow{\partial}_{g}^{+}.
\end{equation}
The left odd derivative $\overrightarrow{\partial}_{g_{i}}^{-}$ has
the following properties:
\begin{equation}
\overrightarrow{\partial}_{g_{i}}^{-}c=0,\,\,\,\overrightarrow{\partial}_{g_{i}}^{-}g_{j}^{-}=\delta_{ij},\,\,\,\overrightarrow{\partial}_{g_{i}}^{-}g_{j}^{+}=0.
\end{equation}
Compound objects are differentiated according to the antilinearity
\begin{equation}
\overrightarrow{\partial}_{g}^{-}\left\{ c_{1}f_{1}\left(g\right)+c_{2}f_{2}\left(g\right)\right\} =\overline{c}_{1}\overrightarrow{\partial}_{g}^{-}f_{1}\left(g\right)+\overline{c}_{2}\overrightarrow{\partial}_{g}^{-}f_{2}\left(g\right),
\end{equation}
 and the anticommutaion relation

\begin{equation}
\overrightarrow{\partial}_{g}^{-}f=\left(\overrightarrow{\partial}_{g}^{-}f\right)+\overline{f}\overrightarrow{\partial}_{g}^{-}.
\end{equation}
Here, for each Grassmann number $g$ we have introduced its involution
$\overline{g}$ as negation of its odd part:
\begin{equation}
\overline{g}=g^{+}-g^{-}.
\end{equation}
The properties of the right derivatives are obtained through complex
conjugation, according to the following relations: 
\begin{equation}
\left[\overrightarrow{\partial}_{g}^{\pm}\right]^{*}=\overleftarrow{\partial}_{g^{*}}^{\pm},\,\,\,\,\left[\overleftarrow{\partial}_{g}^{\pm}\right]^{*}=\overrightarrow{\partial}_{g^{*}}^{\pm}.
\end{equation}
Different derivatives have the following commutation relations:
\begin{equation}
\overrightarrow{\partial}_{g_{i}}^{+}\overrightarrow{\partial}_{g_{j}}^{\pm}=\overrightarrow{\partial}_{g_{j}}^{\pm}\overrightarrow{\partial}_{g_{i}}^{+},\,\,\,\,\overrightarrow{\partial}_{g_{i}}^{-}\overrightarrow{\partial}_{g_{j}}^{\pm}=\left(\pm1\right)\overrightarrow{\partial}_{g_{j}}^{\pm}\overrightarrow{\partial}_{g_{i}}^{-}.
\end{equation}
Left and right derivatives are related as:
\begin{equation}
\overrightarrow{\partial}_{g}^{+}f\left(g\right)=f\left(g\right)\overleftarrow{\partial}_{g}^{+},\,\,\,\overrightarrow{\partial}_{g}^{-}f\left(g\right)=-\overline{f}\left(g\right)\overleftarrow{\partial}_{g}^{-}.
\end{equation}
There is relation between the ordinary calculus derivatives and the
metric Grassmann derivatives:
\begin{multline}
\sum_{\boldsymbol{i}_{n}}\Delta_{A}\left(\boldsymbol{i}_{n}\right)\partial_{G_{A}\left(\boldsymbol{i}_{n}\right)}f\left(g\right)=\left\{ \delta^{+}\overrightarrow{\partial}_{g}^{+}+\delta^{-}\overrightarrow{\partial}_{g}^{-}\right\} f\left(g\right)\\
=f\left(g\right)\left\{ \overleftarrow{\partial}_{g}^{+}\delta^{+}+\overleftarrow{\partial}_{g}^{-}\delta^{-}\right\} .
\end{multline}

\subsection{Integration over grassmann algebra}

In order to work with classical probability we need to integrate it
over Grassmann numbers. Therefore, we introduce integration in the
space of $n$-point functions
\begin{equation}
\int dG_{A}P\left(G_{A}\right)\coloneqq\prod_{i=1}^{M}\prod_{\boldsymbol{i}_{n}}\intop_{\mathbb{C}}dG_{A}\left(\boldsymbol{i}_{n}\right)dG_{A}^{*}\left(\boldsymbol{i}_{n}\right)P\left(G_{A}\right),
\end{equation}
where $\prod_{\boldsymbol{i}_{n}}$ means the product over all the
ordered sequences $i_{1}<\ldots<i_{n}$. Using our definitions, it
can be shown that there is the following integration by parts formula
\begin{multline}
\int dG_{A}f\left(g\right)\left\{ \sum_{\boldsymbol{i}_{n}}\partial_{G_{A}\left(\boldsymbol{i}_{n}\right)}H\left(\boldsymbol{i}_{n}\right)\right\} P\left(G_{A}\right)\\
=-\int dG_{A}P\left(G_{A}\right)\left\{ h^{+}\overrightarrow{\partial}_{g}^{+}+h^{-}\overrightarrow{\partial}_{g}^{-}\right\} f\left(g\right)\\
-\int dG_{A}f\left(g\right)\left\{ \overleftarrow{\partial}_{g}^{+}h^{+}+\overleftarrow{\partial}_{g}^{-}h^{-}\right\} P\left(G_{A}\right).\label{eq:mixed_integration_by_parts}
\end{multline}
From now on we assume that $n$-point functions are always antisymmetric,
and the subscript $A$ will be omitted. 

\section{STOCHASTIC GRASSMANN $B$ REPRESENTATION\label{sec:STOCHASTIC-GRASSMANN-B-REPRESENTATION}}

\subsection{Definition}

Now we are ready to introduce the stochastic interpretation of formal
Grassmann $B$ representation master equation (\ref{eq:B_representation_master_equation}).
The idea is that we introduce random $g$ number vectors $\boldsymbol{g}$
and $\boldsymbol{g}^{\prime}$. The coherent state dyadic is considered
to be a function of these vectors, $\left|\boldsymbol{g}\right\rangle \left\langle \boldsymbol{g}^{\prime*}\right|$.
These coherent states have the following properties: 
\begin{equation}
\widehat{a}_{i}^{\dagger}\left|\boldsymbol{g}\right\rangle =-\overrightarrow{\partial}_{g_{i}}^{\pm}\left|\boldsymbol{g}\right\rangle =\left(\mp1\right)\left|\boldsymbol{g}\right\rangle \overleftarrow{\partial}_{g_{i}}^{\pm},\label{eq:coherent_state_correondence_1-1}
\end{equation}
\begin{equation}
\widehat{a}_{i}\left|\boldsymbol{g}\right\rangle =\widehat{a}_{i}\left(1-g_{p}\widehat{a}_{p}^{\dagger}\right)\left|0\right\rangle =-\overline{g}_{p}\widehat{a}_{i}\widehat{a}_{p}^{\dagger}\left|0\right\rangle =-\overline{g}_{i}\left|0\right\rangle .
\end{equation}
The last equation is problematic: its form is not suitable for construction
of a phase-space representation. However, if $\boldsymbol{g}$ is
odd, so that $g_{p}=g_{p}^{-}$ , then we obtain 
\begin{equation}
\widehat{a}_{i}\left|\boldsymbol{g}^{-}\right\rangle =g_{i}^{-}\left|0\right\rangle =g_{i}^{-}\left(1-g_{p}^{-}\widehat{a}_{p}^{\dagger}\right)\left|0\right\rangle =g_{i}^{-}\left|\boldsymbol{g}^{-}\right\rangle .\label{eq:coherent_state_correspondence_2-1}
\end{equation}
We see that suitable differential correspondences are realized only
when $\boldsymbol{g}$ belongs to the odd sector. Therefore, from
now on we impose this restriction on $\boldsymbol{g}$ and $\boldsymbol{g}^{\prime}$.
The conjugated relations are:
\begin{equation}
\left\langle \left(\boldsymbol{g}^{-}\right)^{*}\right|\widehat{a}_{i}=-\left\langle \left(\boldsymbol{g}^{-}\right)^{*}\right|\overleftarrow{\partial}_{g_{i}^{*}}^{-}=\overrightarrow{\partial}_{g_{i}^{*}}^{-}\left\langle \left(\boldsymbol{g}^{-}\right)^{*}\right|,\label{eq:coherent_state_correspondence_1_conjugated-1}
\end{equation}
\begin{equation}
\left\langle \left(\boldsymbol{g}^{-}\right)^{*}\right|\widehat{a}_{i}^{\dagger}=\left\langle \left(\boldsymbol{g}^{-}\right)^{*}\right|\left(g_{i}^{-}\right)^{*}.\label{eq:coherent_state_correspondence_2_conjugated-1}
\end{equation}
At the time moment $t=0$, the random vectors $\boldsymbol{g}$ and
$\boldsymbol{g}^{\prime}$ should coincide with the vectors of basis
elements, $\boldsymbol{g}=\boldsymbol{e}$ and $\boldsymbol{g}^{\prime}=\boldsymbol{e}^{\prime}$.
However, at later time they begin to diffuse. We express this fact
by inserting integration over probability distribution into grassmann
$B$ representation (\ref{eq:B_representation}): 
\begin{multline}
\widehat{\rho}\left(t\right)=\intop_{\textrm{odd}}d\boldsymbol{G}d\boldsymbol{G}^{\prime*}P\left(\boldsymbol{G},\boldsymbol{G}^{\prime*};t\right)\\
\times\int de_{1}^{\prime*}\ldots de_{M}^{\prime*}de_{M}\ldots de_{1}B\left(\boldsymbol{e},\boldsymbol{e}^{\prime*}\right)\left|\boldsymbol{g}\right\rangle \left\langle \boldsymbol{g}^{\prime*}\right|,\label{eq:stochastic_B_representation}
\end{multline}
with the initial condition
\begin{equation}
P\left(\boldsymbol{G},\boldsymbol{G}^{\prime*};0\right)=\delta\left(\boldsymbol{G}-\boldsymbol{E}\right)\delta\left(\boldsymbol{G}^{\prime*}-\boldsymbol{E}^{\prime*}\right).
\end{equation}
Here, bold capital letters designate vectors $\boldsymbol{G}=\left(G_{0}\ldots G_{M}\right)$,
$\boldsymbol{E}=\left(E_{0}\ldots E_{M}\right)$ etc.; symbol $G_{j}$
means hierarchy of $n$-point functions for $g_{j}$. In fact, our
Grassmann representation is equivalent to ordinary $c$ number phase
space representation 
\begin{equation}
\widehat{\rho}\left(t\right)=\intop_{\textrm{odd}}d\boldsymbol{G}d\boldsymbol{G}^{\prime*}P\left(\boldsymbol{G},\boldsymbol{G}^{\prime*};t\right)\widehat{\Lambda}\left(\boldsymbol{G},\boldsymbol{G}^{\prime*}\right)\label{eq:g_representation_as_c_representation}
\end{equation}
with the overcomplete operator basis
\begin{multline}
\widehat{\Lambda}\left(\boldsymbol{G},\boldsymbol{G}^{\prime*}\right)=\\
\int de_{1}^{\prime*}\ldots de_{M}^{\prime*}de_{M}\ldots de_{1}B\left(\boldsymbol{e},\boldsymbol{e}^{\prime*}\right)\left|\boldsymbol{g}\right\rangle \left\langle \boldsymbol{g}^{\prime*}\right|.
\end{multline}
We call this representation ``stochastic Grassmann $B$ representation''.

\subsection{Equation of motion}

We denote symbolically the relation (\ref{eq:g_representation_as_c_representation})
as
\begin{equation}
P\left(\boldsymbol{G},\boldsymbol{G}^{\prime*};t\right)=\left\{ \widehat{\rho}\left(t\right)\right\} _{P}\left(\boldsymbol{G},\boldsymbol{G}^{\prime*}\right).
\end{equation}
In order to construct master equation for stochastic grassmann $B$
representation, we procceed analogously to section \ref{subsec:grassmann_B_equation_of_motion}:
we find expressions for $\left\{ \widehat{a}_{i}^{\dagger}\widehat{a}_{j}\widehat{\rho}\left(t\right)\right\} _{P}$
etc. Note that since integration by parts formula (\ref{eq:mixed_integration_by_parts})
contains only the combinations $h^{-}\overrightarrow{\partial}_{g}^{-}$
and $\overleftarrow{\partial}_{g}^{-}h^{-}$, there is no rules for
non-conserving terms like $\left\{ \widehat{a}_{j}\widehat{\rho}\left(t\right)\right\} _{P}$.
Using the coherent state properties (\ref{eq:coherent_state_correondence_1-1}),
(\ref{eq:coherent_state_correspondence_2-1}), (\ref{eq:coherent_state_correspondence_1_conjugated-1}),
and (\ref{eq:coherent_state_correspondence_2_conjugated-1}), we find:
\begin{equation}
\widehat{a}_{i}^{\dagger}\widehat{a}_{j}\left|\boldsymbol{g}\right\rangle =g_{j}\overrightarrow{\partial}_{g_{i}}^{-}\left|\boldsymbol{g}\right\rangle .
\end{equation}
Using this relation and its conjugated variant in stochastic Grassmann
$B$ representation (\ref{eq:stochastic_B_representation}) , and
integrating by parts according to (\ref{eq:mixed_integration_by_parts}),
we find:
\begin{equation}
\left\{ \widehat{a}_{i}^{\dagger}\widehat{a}_{j}\widehat{\rho}\left(t\right)\right\} _{P}=-\sum_{\boldsymbol{i}_{n}}\partial_{G_{i}\left(\boldsymbol{i}_{n}\right)}G_{j}\left(\boldsymbol{i}_{n}\right)\left\{ \widehat{\rho}\left(t\right)\right\} _{P},\label{eq:stochastic_B_star_product_mapping_1}
\end{equation}
\begin{equation}
\left\{ \widehat{\rho}\left(t\right)\widehat{a}_{i}^{\dagger}\widehat{a}_{j}\right\} _{P}=-\sum_{\boldsymbol{i}_{n}}\partial_{G_{j}^{\prime*}\left(\boldsymbol{i}_{n}\right)}G_{i}^{\prime*}\left(\boldsymbol{i}_{n}\right)\left\{ \widehat{\rho}\left(t\right)\right\} _{P}.\label{eq:stochastic_B_star_product_mapping_2}
\end{equation}
Representation for quartic terms like $\left\{ \widehat{a}_{i}^{\dagger}\widehat{a}_{j}^{\dagger}\widehat{a}_{k}\widehat{a}_{l}\widehat{\rho}\left(t\right)\right\} _{P}$
can be found by repeated application of Eqs. (\ref{eq:stochastic_B_star_product_mapping_1})-(\ref{eq:stochastic_B_star_product_mapping_2})
and by using the anticommutation relation
\begin{equation}
\overrightarrow{\partial}_{g_{p}}^{-}g_{s}=\delta_{ps}-g_{s}\overrightarrow{\partial}_{g_{p}}^{-}.
\end{equation}
In the stochastic Grassmann $B$ representation, von Neumann equation
(\ref{eq:von_neumann_equation-1}) assumes form:\begin{widetext}
\begin{multline}
\partial_{t}\left\{ \widehat{\rho}\left(t\right)\right\} _{P}=\left\{ \partial_{G_{p}\left(\boldsymbol{i}_{n}\right)}\left(iT_{pq}G_{q}\left(\boldsymbol{i}_{n}\right)-\frac{i}{4}V_{lpql}G_{q}\left(\boldsymbol{i}_{n}\right)\right)-\frac{i}{4}\partial_{G_{p}\left(\boldsymbol{i}_{m}\right)}G_{r}\left(\boldsymbol{i}_{m}\right)\partial_{G_{q}\left(\boldsymbol{i}_{n}\right)}G_{s}\left(\boldsymbol{i}_{n}\right)V_{pqrs}\right.\\
\left.+\left[\partial_{G_{p}^{\prime}\left(\boldsymbol{i}_{n}\right)}\left(iT_{pq}G_{q}^{\prime}\left(\boldsymbol{i}_{n}\right)-\frac{i}{4}V_{lpql}G_{q}^{\prime}\left(\boldsymbol{i}_{n}\right)\right)-\frac{i}{4}\partial_{G_{p}^{\prime}\left(\boldsymbol{i}_{m}\right)}G_{r}^{\prime}\left(\boldsymbol{i}_{m}\right)\partial_{G_{q}^{\prime}\left(\boldsymbol{i}_{n}\right)}G_{s}^{\prime}\left(\boldsymbol{i}_{n}\right)V_{pqrs}\right]^{*}\right\} \left\{ \widehat{\rho}\left(t\right)\right\} _{P}.\label{eq:bare_stochastic_B_master_equation}
\end{multline}
\end{widetext}We see that the evolution equation for the distribution
$\left\{ \widehat{\rho}\left(t\right)\right\} _{P}$ has the form
of Fokker-Planck equation in Stratonovich form \citep{Gardiner2009},
except that it is lacking a number of complex conjugated terms of
the form (see Appendix B of Ref. \citealp{Polyakov2015})
\begin{equation}
\partial_{G_{p}^{*}\left(\boldsymbol{i}_{n}\right)}\left\{ \ldots\right\} +\partial_{G_{p}^{\prime}\left(\boldsymbol{i}_{n}\right)}\left\{ \ldots\right\} .\label{eq:missing_terms-1}
\end{equation}
However, since the Grassmann coherent state dyadic is analytic,
\begin{equation}
\partial_{G_{p}^{*}\left(\boldsymbol{i}_{n}\right)}\left|\boldsymbol{g}\right\rangle \left\langle \boldsymbol{g}^{\prime*}\right|=0,\,\,\,\,\partial_{G_{p}^{\prime}\left(\boldsymbol{i}_{n}\right)}\left|\boldsymbol{g}\right\rangle \left\langle \boldsymbol{g}^{\prime*}\right|=0,
\end{equation}
we can add the required terms to the right hand side of Eq. (\ref{eq:bare_stochastic_B_master_equation})
(see Appendix B of Ref. \citealp{Polyakov2015}). After performing
this addition, we conclude that $\left\{ \widehat{\rho}\left(t\right)\right\} _{P}$
is a joint probability distribution for the stochastic process (in
a Stratonovich sense)
\begin{multline}
dG_{p}\left(\boldsymbol{i}_{n}\right)=-i\sum_{q}T_{pq}G_{q}\left(\boldsymbol{i}_{n}\right)dt+\frac{i}{4}\sum_{lq}V_{lpql}G_{q}\left(\boldsymbol{i}_{n}\right)dt\\
+\sqrt{\frac{\omega_{\gamma}}{2i}}\sum_{\gamma q}O_{pq}^{\left(\gamma\right)}G_{q}\left(\boldsymbol{i}_{n}\right)dX_{\gamma},\label{eq:sde_for_Cn-1}
\end{multline}
\begin{multline}
dG_{p}^{\prime}\left(\boldsymbol{i}_{n}\right)=-i\sum_{q}T_{pq}G_{q}^{\prime}\left(\boldsymbol{i}_{n}\right)dt+\frac{i}{4}\sum_{lq}V_{lpql}G_{q}^{\prime}\left(\boldsymbol{i}_{n}\right)dt\\
+\sqrt{\frac{\omega_{\gamma}}{2i}}\sum_{\gamma q}O_{pq}^{\left(\gamma\right)}G_{q}^{\prime}\left(\boldsymbol{i}_{n}\right)dY_{\gamma}.\label{eq:sde_for_Dn-2}
\end{multline}
Here we have decomposed the pair potential as \citep{Juillet2002,Tessieri2005,Polyakov2015}
\begin{equation}
V_{pqrs}=\sum_{\gamma}\omega_{\gamma}O_{pr}^{\left(\gamma\right)}O_{qs}^{\left(\gamma\right)}.
\end{equation}
The real Wiener increments $dX_{\gamma}$ and $dY_{\gamma}$ obey
to the standard statistics
\begin{equation}
\textrm{E}\left[dX_{\gamma}\right]=\textrm{E}\left[dY_{\gamma}\right]=\textrm{E}\left[dX_{\gamma}dY_{\mu}\right]=0,
\end{equation}
\begin{equation}
\textrm{E}\left[dX_{\gamma}dX_{\mu}\right]=\textrm{E}\left[dY_{\gamma}dY_{\mu}\right]=\delta_{\gamma\mu}.
\end{equation}
We note that Eqs. (\ref{eq:sde_for_Cn-1}) and (\ref{eq:sde_for_Dn-2})
actually form a set of equations for each of the $n$-point functions
$G_{p}\left(i_{1}\ldots i_{n}\right)$ and $G_{p}^{\prime}\left(i_{1}\ldots i_{n}\right)$,
which are uncopled for different $n$ and even for different values
of $i_{1}\ldots i_{n}$. We can multiply each equation for $G_{p}\left(i_{1}\ldots i_{n}\right)$
and $G_{p}^{\prime}\left(i_{1}\ldots i_{n}\right)$ by $e_{i_{1}}\ldots e_{i_{n}}$
and $e_{i_{1}}^{\prime}\ldots e_{i_{n}}^{\prime}$, then sum them
up over $i_{1}\ldots i_{n}$ and over $n$, and obtain a system of
coupled stochastic equations for odd grassmann numbers $g_{1}\ldots g_{M}$
and $g_{1}^{\prime}\ldots g_{M}^{\prime}$:
\begin{multline*}
dg_{p}=-i\sum_{q}T_{pq}g_{q}dt+\frac{i}{4}\sum_{lq}V_{lpql}g_{q}dt\\
+\sqrt{\frac{\omega_{\gamma}}{2i}}\sum_{\gamma q}O_{pq}^{\left(\gamma\right)}g_{q}dX_{\gamma},
\end{multline*}
\begin{multline}
dg_{p}^{\prime}=-i\sum_{q}T_{pq}g_{q}^{\prime}dt+\frac{i}{4}\sum_{lq}V_{lpql}g_{q}^{\prime}dt\\
+\sqrt{\frac{\omega_{\gamma}}{2i}}\sum_{\gamma q}O_{pq}^{\left(\gamma\right)}g_{q}^{\prime}dY_{\gamma}.\label{eq:sde_for_Dn-1-1}
\end{multline}
In fact, the equations are the same as those obtained in \citep{Dalton2016},
except the notational difference for Hamiltonian terms (\ref{eq:standard_hamiltonian-1})
and that our equations are in Stratonovich form, whereas equations
in \citep{Dalton2015} are in Ito form. However, for numerical calculations
we always have to interpret these equations in the $n$-point picture
{[}Eqs. (\ref{eq:sde_for_Cn-1}) and (\ref{eq:sde_for_Dn-2}){]}. 

\section{CONCLUSIONS\label{sec:CONCLUSION}}

A few conclusions can be drawn from the results of this study.

Grassmann numbers are objects of high computational complexity but
they are not as abstract as they are usually considered. We can interpret
$g$ number as a physical many-body state, with a hierachy of correlations.
This leads to natural notions of size and proximity between them.
With the help of these notions, we were able to develop a $c$-number
stochastic calculus on Grassmann algebra. 

Each $g$-number phase-space representation can be converted into
$c$-number phase-space representation by introducing probability
distributions on Grassmann algebra. This way many of $g$-number methods
can be made accessible to computations. 

We put forwad a conjecture that whatever abstract albegraic \textit{one-time
}representation of quantum mechanics is invented, with respect to
classical computability we can always reformulate it as a $c$-number
phase-space representation. This opens up the road for general results
to be established.

We believe that the methods introduced in this work will be useful
when considering such problems as interpretation and $c$-number stochastic
unraveling of (markovian or non-markovian) master equations for open
systems in fermionic environment.
\begin{acknowledgments}
The author acknowledges useful discussions with A.N. Rubtsov. The
study was founded by the RSF, grant 16-42-01057.
\end{acknowledgments}

\bibliographystyle{apsrev4-1}

\end{document}